\documentclass[twocolumn]{aastex63}
\usepackage{amsmath}
\usepackage{textcomp}
\usepackage{amssymb}
\usepackage{natbib}
\usepackage{float}
\usepackage{color}
\usepackage{graphicx}
\usepackage{epstopdf}
\newcommand{\lum}{erg\,s$^{-1}$}
\newcommand{\fermi}{{\it Fermi}}

\newcommand{\chandra}{{\it Chandra}}

\newcommand{\ergflux}{\mbox{${\rm \, erg \,\, cm^{-2} \, s^{-1}}$}}

\newcommand{\gm}{$\gamma$}

\newcommand{\txs}{TXS 2116$-$077}

\shorttitle{Host Galaxy of TXS 2116$-$077}
\shortauthors{Paliya et al.}

\begin{document}

\title{TXS 2116$-$077: A Gamma-ray Emitting Relativistic Jet Hosted in a Galaxy Merger}

\correspondingauthor{Vaidehi S. Paliya}
\email{vaidehi.s.paliya@gmail.com}

\author[0000-0001-7774-5308]{Vaidehi S. Paliya}
\affiliation{Deutsches Elektronen Synchrotron DESY, Platanenallee 6, 15738 Zeuthen, Germany}

\author[0000-0001-9737-4559]{Enrique P{\'e}rez}
\affiliation{Instituto de Astrof\'{i}sica de Andaluc\'ia (CSIC), 18008, Granada, Spain}

\author[0000-0002-7077-308X]{Rub\'en Garc\'ia-Benito}
\affiliation{Instituto de Astrof\'{i}sica de Andaluc\'ia (CSIC), 18008, Granada, Spain}

\author[0000-0002-6584-1703]{Marco Ajello}
\affiliation{Department of Physics and Astronomy, Clemson University, Kinard Lab of Physics, Clemson, SC 29634-0978, USA}

\author{Francisco Prada}
\affiliation{Instituto de Astrof\'{i}sica de Andaluc\'ia (CSIC), 18008, Granada, Spain}

\author[0000-0002-9371-1033]{Antxon Alberdi}
\affiliation{Instituto de Astrof\'{i}sica de Andaluc\'ia (CSIC), 18008, Granada, Spain}

\author[0000-0002-2536-1633]{Hyewon Suh}
\affiliation{Subaru Telescope, National Astronomical Observatory of Japan (NAOJ), 650 North A’ohoku place, Hilo, HI 96720, USA}
\altaffiliation{Subaru Fellow}

\author{C. H. Ishwara Chandra}
\affiliation{National Centre for Radio Astrophysics, TIFR, Post Bag 3, Ganeshkhind, Pune 411007, India}

\author[0000-0002-3433-4610]{Alberto Dom{\'{\i}}nguez}
\affiliation{IPARCOS and Department of EMFTEL, Universidad Complutense de Madrid, E-28040 Madrid, Spain}

\author[0000-0001-5544-0749]{Stefano Marchesi}
\affiliation{INAF - Osservatorio di Astrofisica e Scienza dello Spazio di Bologna, Via Piero Gobetti, 93/3, 40129, Bologna, Italy}

\author{Tiziana Di Matteo}
\affiliation{McWilliams Center for Cosmology, Physics Department, Carnegie Mellon University, Pittsburgh, PA 15213, USA}

\author[0000-0002-8028-0991]{Dieter Hartmann}
\affiliation{Department of Physics and Astronomy, Clemson University, Kinard Lab of Physics, Clemson, SC 29634-0978, USA}

\author[0000-0003-1564-3802]{Marco Chiaberge}
\affiliation{Space Telescope Science Institute, 3700 San Martin Drive, Baltimore, MD 21218, USA}
\affiliation{Johns Hopkins University$-$Center for Astrophysical Sciences, 3400 N. Charles Street, Baltimore, MD 21218, USA}

\begin{abstract}

What triggers collimated relativistic outflows or jets, from the centers of galaxies remains a fundamental question in astrophysics. The merging of two galaxies has been proposed to realize the conditions to successfully launch and drive such jets into the intergalactic medium. However, evidences for the operation of this mechanism are scarce. Here we report the first unambiguous detection of an ongoing merger of a narrow-line Seyfert 1 galaxy, \txs, hosting a closely aligned, \gm-ray emitting relativistic jet with a Seyfert 2 galaxy at a separation of $\sim$12 kpc, using the observations taken with 8.2 m Subaru telescope. Our subsequent followup observations with 10.4 m Gran Telescopio Canarias, 4.2 m William Herschel Telescope, and \chandra~X-ray observatory have provided what is likely to be the first glimpse of the merging environment hosting a closely aligned relativistic jet. Our finding that the jet is considerably younger than the merger demonstrates that jet activity can be triggered by galaxy mergers and that \gm-ray detected narrow-line Seyfert 1 galaxies represent the beginning phase of that activity. These results also highlight the crucial role of mergers in shaping the fate of galaxies in their cosmological evolution and are consistent with recent studies focused on the host galaxy imaging of this enigmatic class of active galactic nuclei.

\end{abstract}


\section{Introduction} \label{sec:intro}
It is now well established that essentially every large galaxy, including our own Milky Way, hosts a massive black hole at its center \citep[e.g.,][]{1995ARA&A..33..581K}. One in a hundred galaxies displays an active galactic nucleus (AGN) where the emission from the nuclear region can outshine the galaxy itself, and controls its growth via strong radiative feedback \citep[cf.][]{2012ARA&A..50..455F}. This emission is due to the black hole accreting circumnuclear gas, but the circumstances that bring the gas near the region of influence of the black hole are still unclear. Among all AGN, $\sim$15\% exhibit spectacular, often large scale, bi-polar relativistic jets \citep[][]{1989AJ.....98.1195K} where particles are confined by the magnetic field and accelerated to near the speed of light. AGN with jets aligned with our line of sight are called blazars and exhibit highly variable emission across the entire electromagnetic spectrum. Blazars are the most powerful, persistent sources of radiation in the universe \citep[e.g.,][]{2014Natur.515..376G,2019ApJ...881..154P}.
 
Jetted AGN, including blazars, are usually hosted in massive, old elliptical galaxies \citep[see, e.g.,][]{2000ApJ...543L.111L,2011MNRAS.416..917C} which are likely to be formed in major merging events \citep[e.g.,][]{1972ApJ...178..623T}. Thus, it is surmised that galaxy mergers are responsible for the launch of relativistic jets \citep[][]{1995ApJ...438...62W,2008ApJS..177..148F,2013MNRAS.436..997R,2015ApJ...806..147C}. However, these studies were focused mostly on quasars with off-axis jets, i.e., sources viewed at large angles with respect to jet axis ($\gtrsim$30$^{\circ}$). Moreover, quasars studied in previous works were found to reside in already evolved ellipticals or exhibiting morphologies as disturbed as those expected from a late stage merger \citep[][]{2015ApJ...806..147C}. In blazars, on the other hand, the on-axis viewing geometry makes study of the host galaxy extremely difficult due to Doppler boosted, bright jet emission and results acquired from various campaigns focusing on the host galaxy properties of blazars were remain inconclusive, albeit providing hints in support of the findings derived by studying off-axis jets  \citep[e.g.,][]{1998A&A...336..479K,2003A&A...400...95N,2016MNRAS.460.3202O}. The missing link, however, is to find examples of closely aligned jet systems caught in the act of merging and/or hosted in late-type (i.e., young) AGN, which can reveal the stage in the AGN evolution when the jet activity was recently triggered.
 
Among all classes of AGN, \gm-ray detected narrow-line Seyfert 1 (\gm-NLSy1) galaxies are probably the optimal systems as they are known to host closely aligned, low-power relativistic jets \citep[][]{2009ApJ...707L.142A,F15,2019ApJ...872..169P} and thus probably the best systems where to explore the jet-host-galaxy connection. NLSy1 galaxies are classified based on their optical spectral characteristics, such as broad permitted lines with small widths (H$\beta$ FWHM $<$2000 km s$^{-1}$), weak [O III] emission ([O III]/H$\beta$ flux ratio $<$3), and strong Fe complexes \citep[][]{1985ApJ...297..166O,1989ApJ...342..224G}. The detection of variable \gm-ray emission from about half-a-dozen radio-loud NLSy1 objects with the \fermi~Large Area Telescope \citep[e.g.,][]{2009ApJ...707L.142A} has provided the strongest evidence in support of the relativistic beaming. These enigmatic sources, in general, are thought to be rapidly accreting \citep[][]{1992ApJS...80..109B}, low-luminosity AGN in the early phase of their evolution \citep[][]{2000MNRAS.314L..17M,2004ApJ...606L..41G,2018A&A...614A..87B}. They have also been found to reside in spiral/late-type galaxies \citep[][]{2016ApJ...832..157K,2019AJ....157...48B,2020MNRAS.492.1450O} providing supportive evidences about their young nature.

The facts that make \gm-NLSy1s as the most promising systems to study the jet launching processes are their relatively low jet powers and young nature \citep[e.g.,][]{F15}. A low jet power indicates that the underlying host galaxy properties can be explored in a much better fashion than that is possible for more powerful blazars. Furthermore, studying a young \gm-NLSy1 also provides an opportunity to reveal the jet-host galaxy interaction when the system is rapidly growing and possibly hosts a nascent jet.

Here we report observations of TXS 2116$-$077, a NLSy1 galaxy \citep[][]{2017ApJS..229...39R} with redshift $z=0.26$ \citep[corresponding to a distance of 1329 Mpc;][]{2016A&A...594A..13P}. TXS 2116$-$077 possesses a relativistic jet closely aligned to our line of sight as confirmed by the detection of its variable \gm-ray flux, the observation of a compact radio core at 8.4 GHz, its large radio-loudness, and flat radio and X-ray spectra  \citep[][]{2018ApJ...853L...2P,2018MNRAS.477.5127Y}. The broadband spectral energy distribution of TXS 2116$-$077 resembles that of blazars \citep[][]{2018MNRAS.477.5127Y,2019ApJ...872..169P}, which again confirms the presence of a closely aligned relativistic jet in this NLSy1 galaxy. Furthermore, the low-resolution Sloan Digital Sky Survey (SDSS) optical image of this source exhibits an extended structure, which was tentatively explained as possibly due to merger or enhanced star-forming activities \citep[][]{2018ApJ...853L...2P,2018MNRAS.477.5127Y}.

We have carried out, for the first time, a high-resolution, adaptive optics enabled, $J$-band imaging of the host galaxy of TXS 2116$-$077 with 8.2 m Subaru telescope, revealing two galaxies in the act of merging. We have followed it up with long-slit optical spectroscopy and  integral-field unit (IFU) observation with 10.4 m Gran Telescopio Canaris (GTC) and 4.2 m William Herschel Telescope (WHT), and a 10 ksec \chandra~X-ray observatory pointing to study the host galaxy environment of a beamed AGN in an unprecedented detail. In Section~\ref{sec:analysis}, we briefly describe the adopted data reduction procedure and present the acquired results in Section~\ref{sec:results}. We discuss our findings in Section~\ref{discussion} and summarize in Section~\ref{summary}.

\section{Data Reduction and Analysis}\label{sec:analysis}
\subsection{Host Galaxy Imaging with Subaru}
We obtained high-resolution $J$-band image of \txs~with the InfraRed Camera and Spectrograph \citep[IRCS;][]{2000SPIE.4008.1056K}, in combination with its adaptive optics (AO) system AO188 \citep[][]{2010SPIE.7736E..0NH}, mounted at the Subaru Telescope on 2018 June 16. We used a resolution of 0.052\arcsec~per pixel for a corresponding 54\arcsec~field of view. The Laser Guide Star mode was used for AO and observations were performed in imaging mode. The total integration time was 34 minutes while accumulating 17 frames with an exposure time of 120 seconds per frame using a 5 point dither pattern. The natural seeing was $\sim$0.7\arcsec~and the AO-corrected seeing was 0.2\arcsec.

The IRCS data was analyzed with Image Reduction and Analysis Facility \citep[{\tt IRAF};][]{1986SPIE..627..733T}. This includes the creation of a bad pixel mask, background subtraction, flat fielding, estimate of the dither offsets, and final stacking of the data. A nearby star, SDSS J211853.33$-$073214.3 (right ascension: 21$^{\rm h}$ 18$^{\rm m}$ 53.33$^{\rm s}$, declination: $-$07$^{\circ}$ 32$^{\prime}$ 14.34$^{\prime\prime}$), located $\sim$14\arcsec~North-East of the \gm-NLSy1 was used as a reference for the flux calibration.

\subsection{Long-Slit Optical Spectroscopy}
{\bf ISIS at WHT:} Long-slit spectrophotometric observations of TXS 2116$-$077 were obtained in service time with the Intermediate-dispersion Spectrograph and Imaging System (ISIS) double-arm spectrograph mounted on WHT of the Isaac Newton Group (ING) at the Roque de los Muchachos Observatory on the Spanish island of La Palma. They were acquired on 2018 August 13 with an average seeing of 0.9\arcsec. To cover both nuclei detected in the Subaru image, we used a slit position angle of 116.7$^{\circ}$.  Grating R158B was used in the blue arm centered at 4500 \AA\ with a wavelength range between 3370$-$5400 \AA\ and 1.45 \AA\ pixel$^{-1}$ dispersion. In the red arm, grating R158R centered in 7500 \AA\ covered between 5300$-$10000 \AA\ with a dispersion of 1.82 \AA\ pixel$^{-1}$. The slit width was 1\farcs2. A total of six exposures of 1800 seconds each were taken simultaneously in each arm. The reduction was performed using {\tt IRAF} routines in the usual manner. For both arms, the standard star G191-B2B was used to produce final sky subtracted and flux calibrated spectra.

{\bf OSIRIS at GTC:} Long-slit spectrophotometric observations were obtained under Director's Discretionary Time (DDT) program using the Optical System for Imaging and low-Intermediate-Resolution Integrated Spectroscopy (OSIRIS) spectrograph mounted at GTC located at the Roque de los Muchachos Observatory on the island of La Palma, Spain. They were taken on August 16, 2018 with an average seeing of 0.9\arcsec~during dark time using grating R2000B and a slit width of 0\farcs8. The wavelength range covered between 3950$-$5650 \AA\ at a dispersion of 0.86 \AA\ pixel$^{-1}$. A total of four exposures of 1800 seconds each were taken. The reduction was performed using {\tt IRAF} routines following the standard procedures. The standard star Ross 640 was used to obtain the sensitivity curve and perform the flux calibration.

\subsection{Integral Field Unit Observations}
Multi-Espectr{\'o}grafo en GTC de Alta Resoluci{\'o}n para Astronom{\'\i}a (MEGARA) is an optical IFU and multi-object spectrograph installed at the GTC \citep[][]{2016SPIE.9908E..1KG}. The MEGARA IFU encompasses 567 contiguous hexagonal fibers, each with a long diagonal of 0\farcs62, resulting in a 12.5$\times$11.3 arcsec$^2$ field of view in the shape of a rectangle. Sky observations are obtained simultaneously with eight fiber bundles with 7 dedicated hexagonal fibers each located at the outermost parts of the field of view (R$>$1.5$^{\prime}$). MEGARA IFU spectra of TXS 2116$-$077 were obtained under DDT program on the night of 2018 October 1 using the LR-I VPH grating that covers from $\sim$7200$-$8650 \AA\ with a reciprocal linear dispersion of 0.37 \AA\ pixel$^{-1}$. We obtained six exposures of 1690 seconds each. The seeing conditions were around 0\farcs6. To flux-calibrate the LR-I spectra, several exposures of the flux standard star HR 7950 were obtained under similar airmass and seeing conditions. We also obtained calibration data, including ThNe and ThAr arc lamps and halogen lamps.

The raw IFU spectra were processed using the MEGARA data reduction pipeline\footnote{https://github.com/guaix-ucm/megaradrp} (MDRP; S. Pascual 2020, in preparation). The main steps include: bias correction, trimming, identification of the position of the spectra on the detector along the dispersion axis, model mapping and extraction of each individual spectrum, fiber-flat correction, wavelength calibration, flux calibration, and sky subtraction. The final products of the MDRP are Row-Stacked Spectra (RSS) 2D images containing 623 fiber spectra. The H$\alpha$ line shows a second, broad component in some spaxels around the Seyfert 1 nucleus. Thus, we used a multi-component Gaussian model to fit the [N II] + H$\alpha$ window. For the sake of visualization, we have transformed the hexagonal spaxel grid of the RSS data into a square pixel shape grid of 0\farcs25 per pixel.

 \begin{figure*}[t!]
{\hspace{1.4cm}
\includegraphics[scale=0.5]{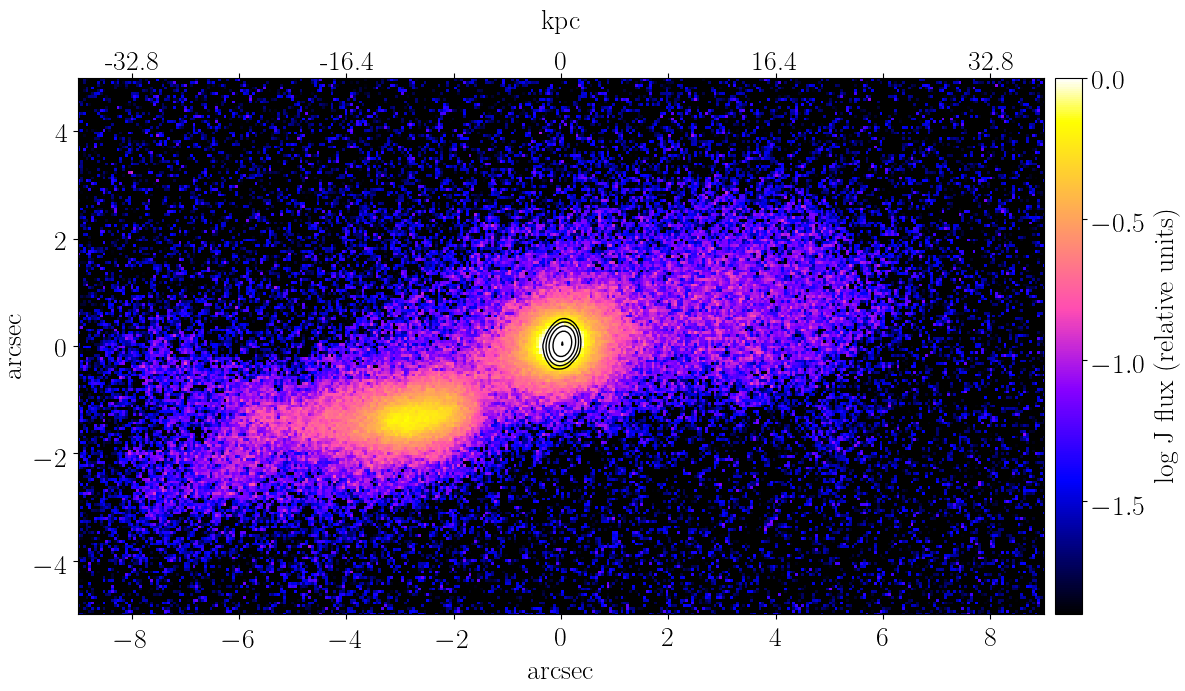}
}
\caption{Adaptive optics enabled $J$-band image of TXS 2116$-$077 ($z=0.26$) obtained with IRCS mounted at the Subaru telescope. The image is centered at the \gm-NLSy1 galaxy. The colorbar represents the relative flux on a logarithmic scale. The image scale is 0.052$^{\prime\prime}$ per pixel. Overplotted are 8.4 GHz VLA contours (levels from 0.2 mJy beam$^{-1}$, i.e., five times the off-source rms, to 89.7 mJy beam$^{-1}$ in five logarithmic intervals). The beam size is 0.32$^{\prime\prime}$ $\times$ 0.23$^{\prime\prime}$ with a position angle of $-$13.81$^{\circ}$. Top axis shows the scale in kpc with 1\arcsec=4.06 kpc. North is up and East to the left.}
\label{subaru}
\end{figure*}
\subsection{Very Large Array Observations}
The archival Very Large Array (VLA) data of TXS 2116$-$077 at 8.4 GHz, observed on 1998 May 18, was analyzed using Astronomical Image Processing System ({\tt AIPS}), following standard procedures. The flux scale was set using the standard flux calibrator 3C 286. The flux and phase calibrations were transferred to the target via the secondary calibrator. A few rounds of self-calibration were performed to correct for phase variations, which improved the image quality. The resolution of the final image is 0.32\arcsec$\times$0.23\arcsec~at a position angle of $-$14 degrees. The root mean square (rms) noise floor of the map is 42 $\mu$Jy/beam. The \gm-NLSy1 remained unresolved at this frequency and resolution. We used the Gaussian fitting program {\tt JMFIT} in {\tt AIPS} to get the flux density. The peak and integrated flux density is nearly identical confirming the compact nature of the source. The integrated flux density of the target is 91.87 mJy. To compute the error on the flux, we have added the calibration error, Gaussian fitting error, and the rms noise of the map in quadrature, giving the final error as 0.17 mJy. The companion Seyfert 2 galaxy, on the other hand, remain undetected and we derived the 3$\sigma$ flux upper limit as 0.13 mJy.

A deep radio imaging of TXS 2116$-$077 with Very Large Baseline Array (VLBA) has recently been performed on 2018 March 30. We analyzed this publicly available VLBA observation at 6 cm (5 GHz). A sustained data-recording rate of 512 Mbits/s in two-bit sampling was used. The frequency band was split into 4 intermediate frequencies (IFs) of 64 MHz bandwidth each, for a total  synthesized bandwidth of 256 MHz. Each IF was in turn split into 256 channels of 250 KHz bandwidth. The data were corrected for Earth-orientation and ionospheric effects. The source BL~Lac was used as bandpass calibrator. The data were correlated at the NRAO data processor using an averaging time of 2 seconds. We performed standard a-priori gain calibration using the measured gains and system temperatures of each antenna. This calibration, as well as the data inspection and editing, were done within {\tt AIPS}. The flux scale was set using the calibrator BL~Lac. The observation resulted in a beam of 5.3 $\times$ 2.4 millarcsec$^2$, at a PA of $-$2 degrees. A compact radio emission positionally consistent with the \gm-NLSy1 galaxy is found with no hints of any extended features or knots. The peak flux maximum is located at the position the Right Ascension: 21$^{\rm h}$ 18$^{\rm m}$ 52.96$^{\rm s}$ and Declination, J2000 = $-$07$^{\circ}$ 32$^{\prime}$ 27.58$^{\prime\prime}$, with a flux density of 27.93 $\pm$ 0.23 mJy/b. There is no detection of the companion Seyfert 2 galaxy, with an upper limit of 0.13 mJy/beam.

\subsection{\textit{Chandra} Observations}
\chandra\ X-ray observatory carried out a DDT observation of TXS 2116$-$077 on 2018 September 15 for a net exposure of 10\,ksec. We used the nominal aim point of the Advanced CCD Imaging Spectrometer chip (ACIS-S3) and the data were collected in a full readout mode resulting in 3.241 seconds readout time. The data were reprocessed using {\tt chandra\_repro} and an exposure-corrected image was generated using {\tt fluximage}. We scanned the \chandra~image with the source detection algorithm {\tt wavedetect} to search for X-ray emitting objects. A bright X-ray source (Right ascension, J2000 = 21$^{\rm h}$ 18$^{\rm m}$ 52.95$^{\rm s}$ and Declination, J2000 = $-$07$^{\circ}$ 32$^{\prime}$ 27.54$^{\prime\prime}$) coincident with the \gm-NLSy1 is identified with the separation between the X-ray and radio positions as 0.19$^{\prime\prime}$. The excellent spatial resolution of  \chandra~($\sim$0.5$^{\prime\prime}$) allowed us to resolve the merging system, however the companion Seyfert 2 galaxy remains below the detection limit.

We used the task {\tt specextract} to extract the source and background spectra and corresponding ancillary and response matrix files. A source region of 1.5$^{\prime\prime}$ centered at the \gm-NLSy1 galaxy was adopted and we selected the background as a circle of 10$^{\prime\prime}$ from a nearby source-free region. We rebinned the source spectrum to have 1 count per bin and performed the fitting in the energy range of 0.5$-$7 keV using C-statistics \citep[][]{1979ApJ...228..939C} in XSPEC \citep[][]{Arnaud96}. 

\section{Results}\label{sec:results}
\begin{figure}[t!]
\vbox{\centering
\includegraphics[width=\linewidth]{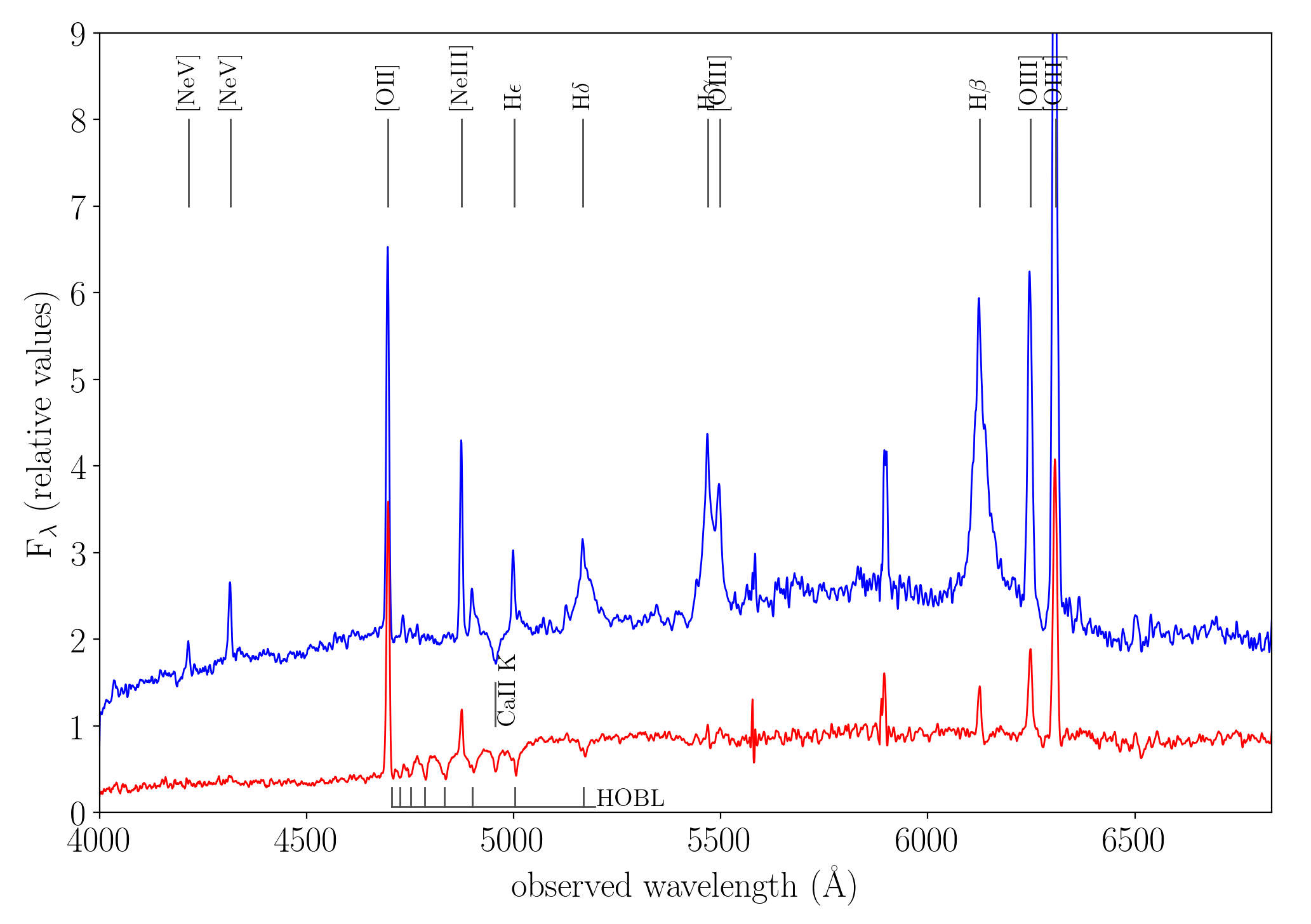}
}
\caption{Integrated long-slit spectra of the two galaxies, the \gm-NLSy1 (blue) and the Seyfert 2 (red), taken with ISIS at WHT and OSIRIS at GTC. The main emission and absorption lines are labeled, including the higher order Balmer lines (HOBL) in the Seyfert 2.}
\label{subaru_spec}
\end{figure}

We show the high-resolution, adaptive optics enabled, $J$-band image of \txs~in Figure~\ref{subaru} and overplot the 8.4 GHz VLA contours which suggests the presence of a compact radio jet associated with the \gm-NLSy1 galaxy. As can be seen, two galactic nuclei are resolved (separation $\sim$12 kpc) and appear to be interacting. This merging phenomenon is confirmed from the long-slit optical spectroscopy which reveals both objects to be located at the same redshift, as shown in Figure~\ref{subaru_spec}. The optical spectrum of the companion galaxy shows emission lines, thus hosts an AGN, which, in the absence of broad emission lines, can be classified as a Seyfert 2.

\begin{figure*}[t!]
\hbox{\hspace{3.0cm}
\includegraphics[scale=0.5]{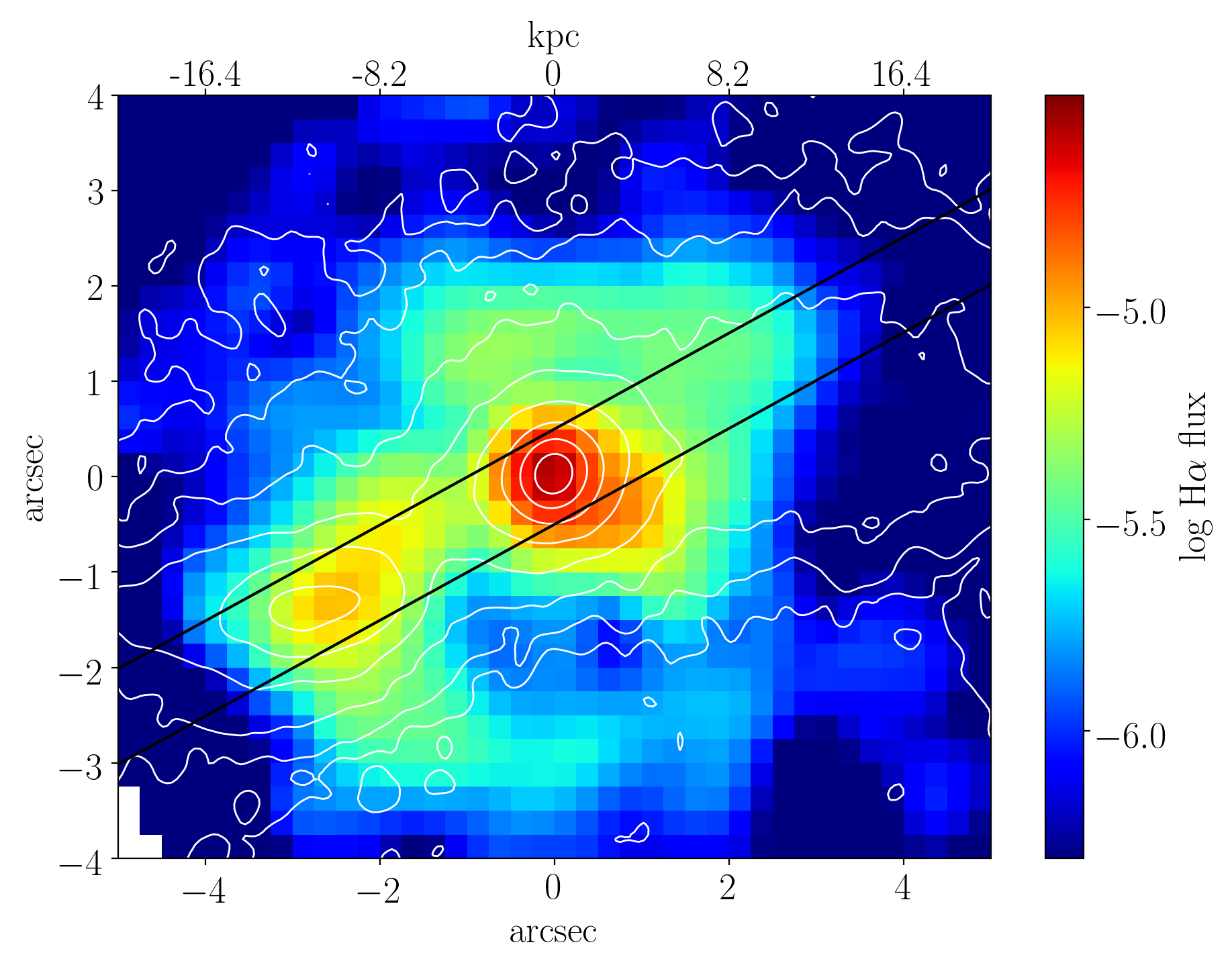} 
}
\hbox{\vspace{0.0cm}
\includegraphics[width=\textwidth]{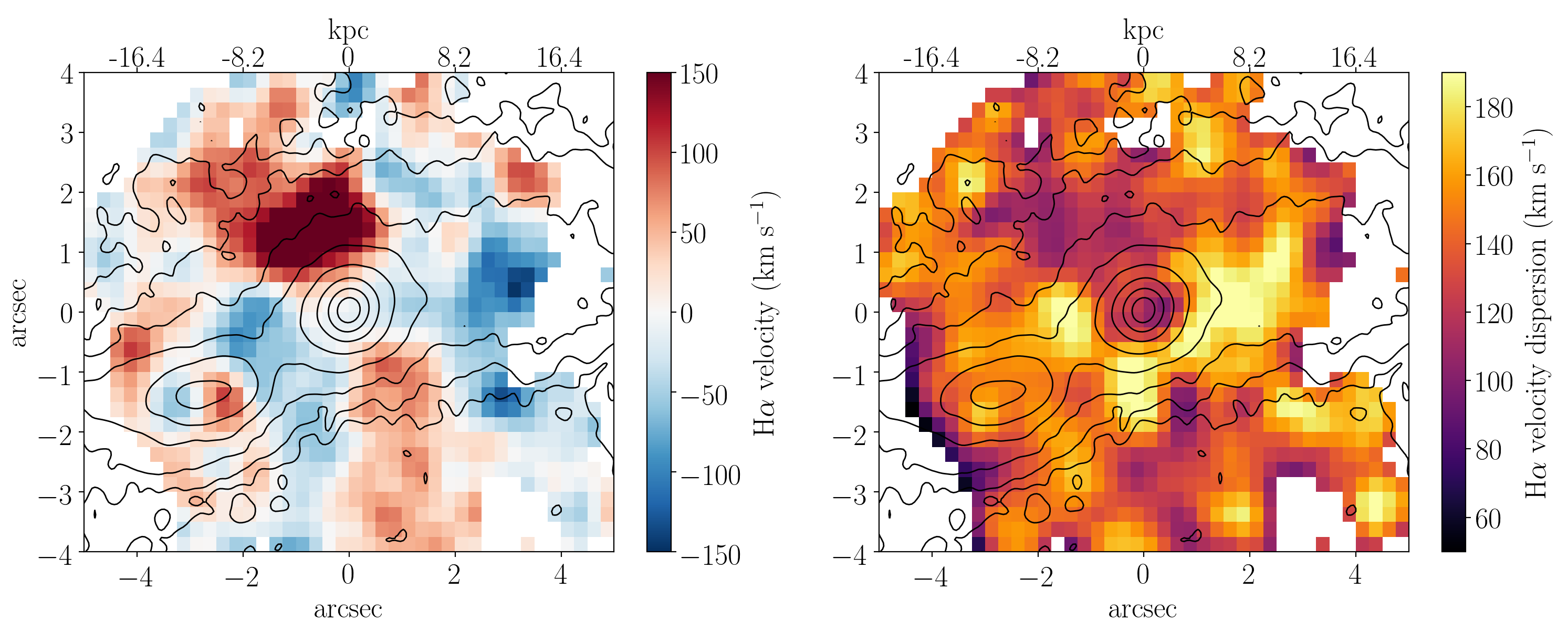} 
}
\caption{Top: The H$\alpha$ flux (only the narrow component, relative flux units in logarithmic scale) derived from the integral field unit MEGARA at GTC. Contours are $J$-band image (Figure~\ref{subaru}) logarithmically scaled in intervals of 0.1 dex starting at 3$\sigma$ background. A representation of the slit position for the long-slit spectrographs ISIS at WHT and OSIRIS at GTC is shown with parallel black lines. Bottom: Spatial distribution of the velocity (left) and velocity dispersion (right).  We have performed a multi-Gaussian fitting to the H$\alpha$ line in the MEGARA IFU data. In this plot, we present the data from the narrow component.
}
\label{sub_J_meg_ha_maps}
\end{figure*}

The H$\alpha$ flux map acquired from the integral field spectroscopy unit MEGARA mounted on the GTC shows two strong peaks, spatially coincident with both nuclei (Figure~\ref{sub_J_meg_ha_maps}). The \gm-NLSy1 nucleus is surrounded by a $\sim$2$^{\prime\prime}$ ring-like distribution of H$\alpha$ emitting material and a curved filament connects both nuclei and extends to the south of the \gm-NLSy1. The H$\alpha$ velocity and velocity dispersion maps as shown in the bottom panel of Figure~\ref{sub_J_meg_ha_maps}, reveal complex kinematic structures within the merging system. There is no clear large-scale systematic rotation pattern, with patchy blue- and red-shifted regions (relative to the \gm-NLSy1 nucleus). This observation is consistent with the merging driven turbulence, possibly due to merger induced shocks, the presence of the relativistic jet, and their interaction. Interestingly, there is a fast outflowing region detected at $\sim$1$^{\prime\prime}$ North-East of the \gm-NLSy1 nucleus, possibly hinting a buried or nascent jet \citep[][]{2010MNRAS.401....7H}.

\begin{figure}[t!]
\includegraphics[width=\linewidth]{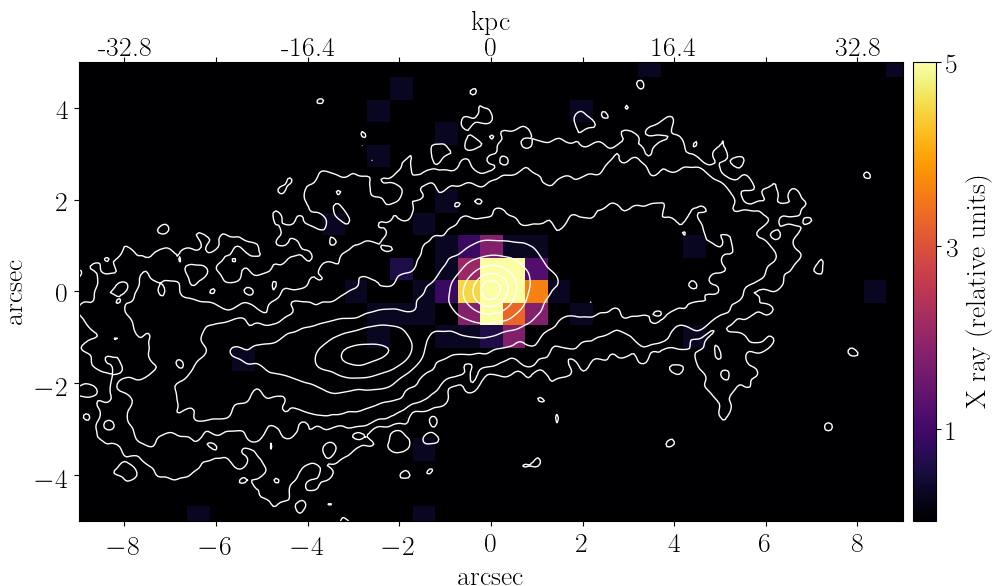}
\caption{The 0.5$-$7 keV, exposure corrected \chandra~X-ray image of the merging system TXS 2116$-$077. The resolution of the image is 0.49\arcsec~per pixel. For a comparison, we overplot the Subaru IR contours as in Figure~\ref{sub_J_meg_ha_maps}. North is up and East to the left.}
\label{chandra}
\end{figure}

The High-resolution, 0.5$-$7 keV image of the merging system \txs~is shown in Figure~\ref{chandra} and we overlaid the Subaru contours to guide the eyes. Only the \gm-NLSy1 galaxy is detected whose spectrum is well-fitted (C-statistics=146.1 for 166 degree of freedom) by a simple power law corrected for Galactic absorption \citep[$N_{\rm H}$=6.99$\times$10$^{20}$ cm$^{-2}$;][]{Kalberla05}. The best-fit photon index is $\Gamma$ = 1.64$^{+0.19}_{-0.18}$, and the 0.5--7\,keV unabsorbed energy flux is  $f_{\rm 0.5-7} = 4.27^{+0.49}_{-0.39}\times10^{-13}$ \ergflux. Such a flat X-ray spectrum provides a supportive evidence of the jet based origin of the observed X-ray emission \citep[see also][]{2019ApJ...872..169P}. Furthermore, the corresponding $K$-corrected, isotropic luminosity is $L_{\rm 0.5-7}=8.71^{+1.03}_{-0.95}\times 10^{43}$ \lum. For the Seyfert 2 galaxy, we derived the 3$\sigma$ flux upper limit, $f_{\rm 0.5-7}$ $<$ 3.98$\times$10$^{-16}$ \ergflux~($L_{\rm 0.5-7}<8.80\times 10^{40}$ \lum), assuming a typical photon index $\Gamma$=2 and a line of sight column density as that of \gm-NLSy1. If the source is a Compton thick AGN (with intrinsic $N_{\rm H}=1\times10^{24}$ cm$^{-2}$), the 3$\sigma$ unabsorbed flux upper limit is instead $f_{\rm 0.5-7}$ $<$ 2.78$\times$10$^{-14}$ \ergflux~($L_{\rm 0.5-7}<6.17\times 10^{42}$ \lum).

\section{Discussion}\label{discussion}
\subsection{Host Galaxy and Merger Environment}
The results acquired from the Subaru imaging and subsequent followup observations of \txs~confirm its ongoing merger with a Seyfert 2 galaxy. This is probably the first unambiguous detection of a \gm-ray emitting jet hosted by a merging system and hints that relativistic jets can be triggered by galaxy mergers \citep[][]{2015ApJ...806..147C}. Note that a few recent studies have also reported the identification of minor merging events associated with NLSy1 galaxies \citep[e.g.,][]{2018A&A...619A..69J,2020MNRAS.492.1450O}, however, due to lack of spectroscopic redshift measurement of the companion, strong claims could not be made. \citet[][]{2019AJ....157...48B} presented the spectroscopic confirmation of the merger of a radio-loud NLSy1 galaxy IRAS 20181$-$2244 ($z=0.185$), however, this objet is yet to be detected in the \gm-ray band and thus its extent of relativistic beaming is uncertain. Furthermore, the small physical separation ($\sim$12 kpc) and the presence of AGN activity in both galaxies indicates that this is a merger in the late stage of evolution, i.e., close to coalescence \citep[][]{2005Natur.433..604D,2012ApJ...748L...7V}. Both galaxies are embedded in a common, 70 kpc extended envelope composed of gas and stars, and the companion galaxy exhibits morphological features associated with tidal disruption (Figure~\ref{subaru}).

The photometric decomposition of the Subaru image was performed with the 2D fitting algorithm GALFIT \citep[][]{2010AJ....139.2097P}. The star SDSS J211853.33$-$073214.3 was used to model the point spread function and remove the AGN contamination. A nearby source-free region was used as the sky background. First, we fitted both galaxies only with the PSF and then added components, mainly S\'{e}rsic profiles, based on the leftover emission in the residual image and the reduced $\chi^2$ value. Typically, a smaller value of the S\'{e}rsic index ($n\lesssim$2) is associated with galaxies with disk-like morphology and pseudobulges, and larger values of $n$ ($>$2) with elliptical galaxies and classical bulges.

\begin{figure*}[t!]
\centering
\includegraphics[width=\textwidth]{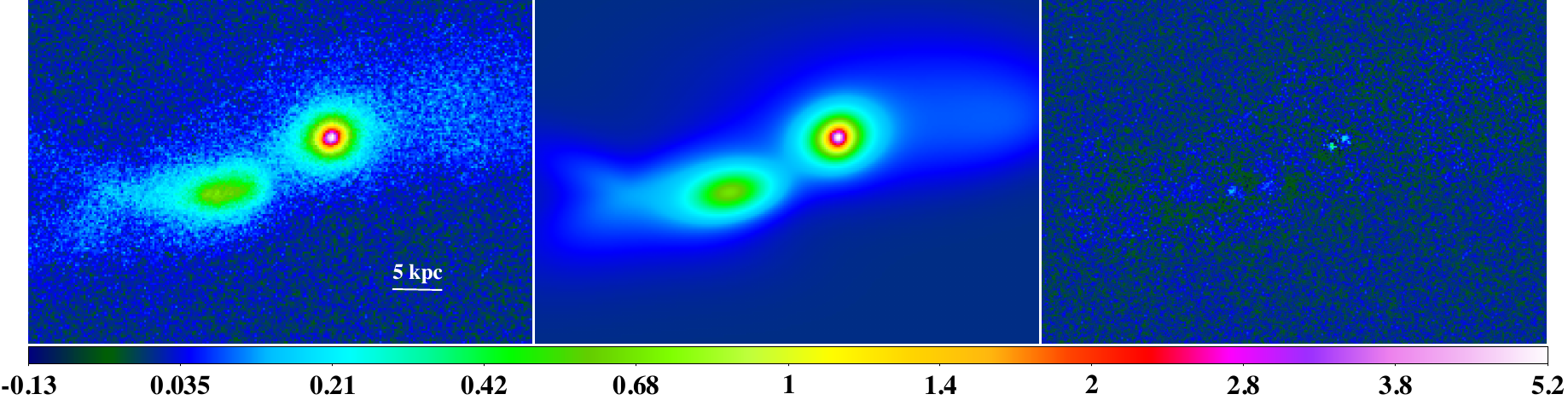}
\caption{The observed Subaru image (left), model image (middle) generated with GALFIT, and the residual of the fit (right). The spatial scale at the object redshift is indicated in the left panel. Other information are same as in Figure~\ref{subaru}.}
\label{galfit}
\end{figure*}

The best-fitted model includes two S\'{e}rsic functions for each AGN and three Fourier components \citep[see][for details]{2010AJ....139.2097P} to take into account the extended emission. The resultant fitting parameters are provided in Table~\ref{tab:galfit} and the results of the fitting are shown in Figure~\ref{galfit}. For both \gm-NLSy1 and the companion, first S\'{e}rsic functions with $n=1.48\pm0.06$ and $0.73\pm0.03$, respectively, resemble that of pseudobulges in late-type galaxies. The physical interpretation of the second S\'{e}rsic components is tedious due to disturbed morphologies of both galaxies. However, the small S\'{e}rsic index for this component indicates the presence of disk/bar in both galaxies. Altogether, it can be concluded that both merging galaxies are `young' and host pseudobulges, similar to what is commonly known for other NLSy1 \citep[][]{2000MNRAS.314L..17M,2018A&A...619A..69J,2020MNRAS.492.1450O}. This reinforces the hypothesis that \gm-NLSy1 galaxies, and NLSy1s in general, are young sources within the AGN evolution scheme.

\begin{deluxetable}{lcc}
\tabletypesize{\small}
\tablecaption{Best-fit GALFIT modeling parameters.\label{tab:galfit}}
\tablewidth{0pt}
\tablehead{
\colhead{} & \colhead{Function} & \colhead{}\\
\colhead{Parameter} & \colhead{S\'{e}rsic 1} & \colhead{S\'{e}rsic 2}}
\startdata
\colhead{} & \colhead{\gm-NLSy1 galaxy} & \colhead{}\\
Mag                        & 19.84$\pm$0.01    & 19.49$\pm$0.01 \\
$r_{\rm e}$ (kpc)  & 2.35$\pm$0.05      & 8.94$\pm$0.24 \\
$n$                        & 1.48$\pm$0.06      & 0.53$\pm$0.01 \\
Axial ratio              & 0.68$\pm$0.01      & 0.52$\pm$0.01 \\
PA ($^{\circ}$)     & $-$79.54$\pm$1.25 & 97.24$\pm$0.51\\
\hline
\colhead{} & \colhead{Seyfert 2 galaxy} & \colhead{}\\
Mag                        & 20.03$\pm$0.01    & 19.91$\pm$0.02 \\
$r_{\rm e}$ (kpc)  & 3.46$\pm$0.03      & 8.47$\pm$0.14 \\
$n$                        & 0.73$\pm$0.03      & 0.71$\pm$0.02 \\
Axial ratio              & 0.44$\pm$0.01      & 0.53$\pm$0.01 \\
PA ($^{\circ}$)     & $-$80.48$\pm$0.46 & $-$88.15$\pm$0.79\\
\enddata
\tablecomments{$n$ and PA are the S\'{e}rsic index and the position angle, respectively. The reduced $\chi^2$ of the fit is 1.16.}
\end{deluxetable}

\begin{figure}[t!]
\centering
\includegraphics[width=\linewidth]{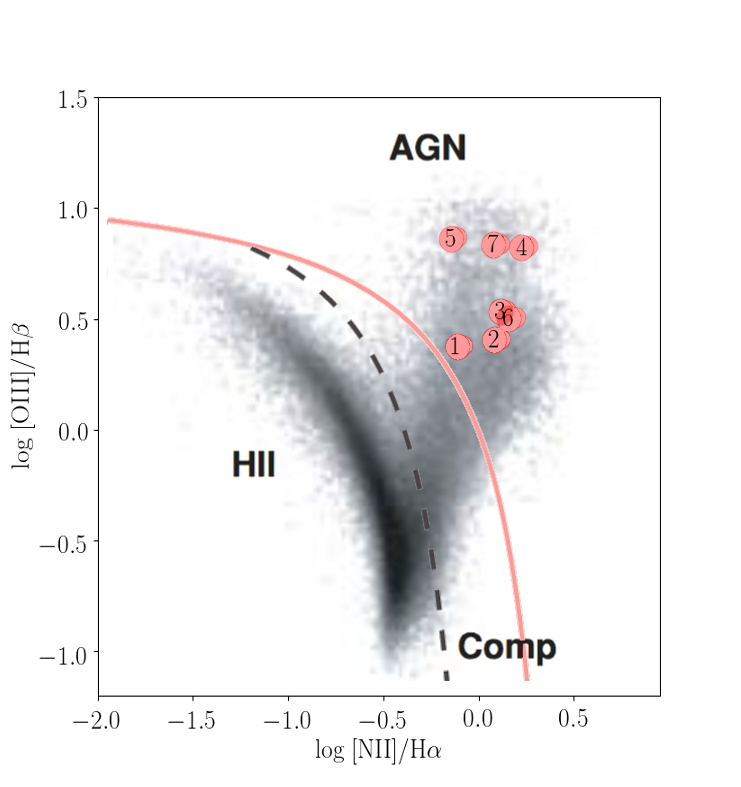} 
\caption{The BPT diagram \citep[][]{1981PASP...93....5B}. Large red dots refer to the individual spectra extracted along the long-slit (see Figure~\ref{sub_J_meg_ha_maps} and \ref{spatial_variations}). Dots corresponding to \gm-NLSy1 and Seyfert 2 nuclei are labeled `5' and `4', respectively. The grey data points refer to 85224 galaxies from the SDSS DR4 \citep[][]{2006MNRAS.372..961K}. The region labeled `HII' marks the sequence of ionization from hot massive stars; whereas, that labeled with `AGN' marks ionization by a harder, AGN non-stellar continuum or by shocks. The two lines drawn separate the pure stellar from the pure AGN ionization in a `Composite' region. Three upper points clearly indicate AGN ionization, while the other four lower points are in a region where both AGN and shock ionization dominate.}\label{bpt}
\end{figure}

\begin{figure*}[t!]
\centering
\includegraphics[scale=0.4]{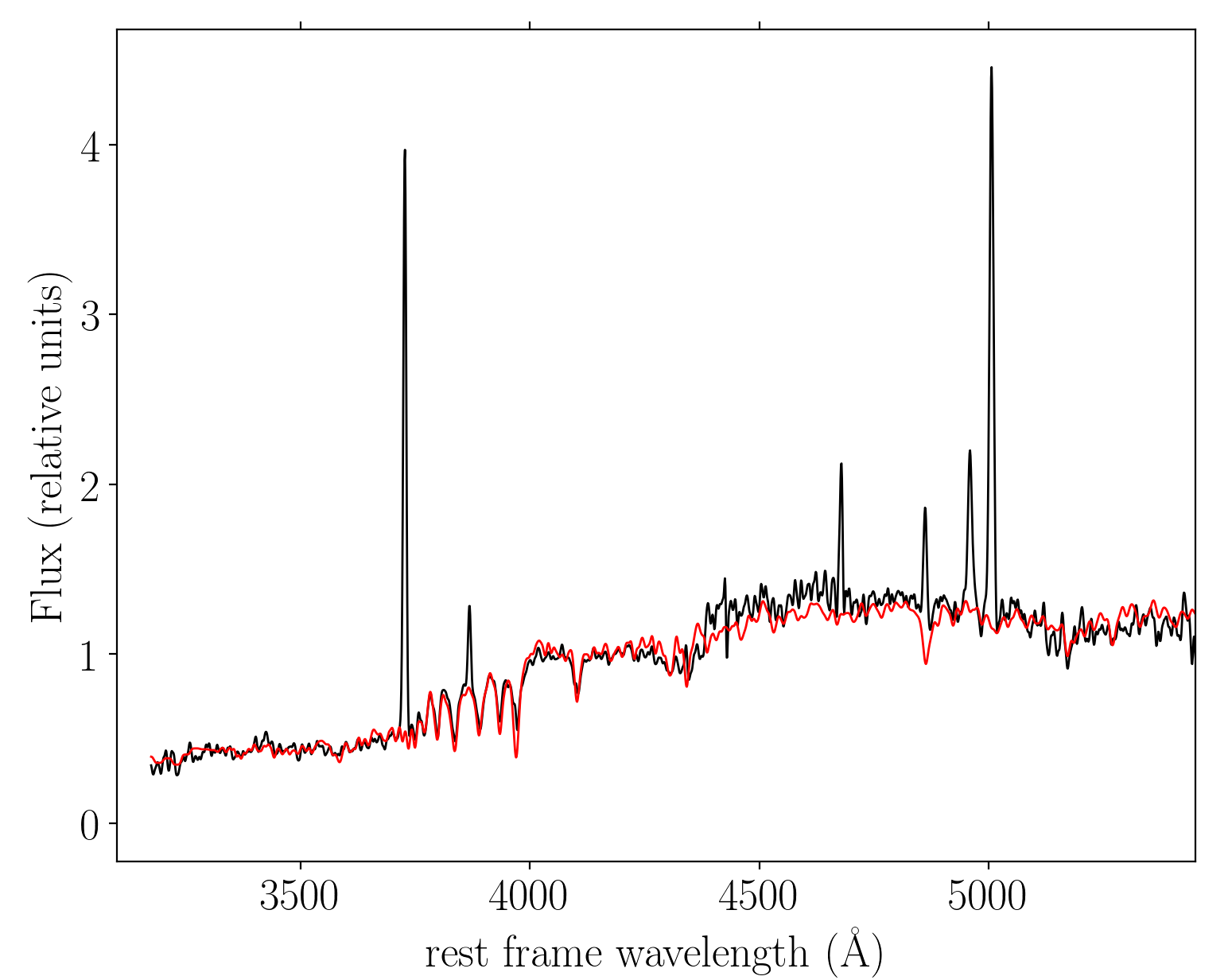} 
\includegraphics[scale=0.5]{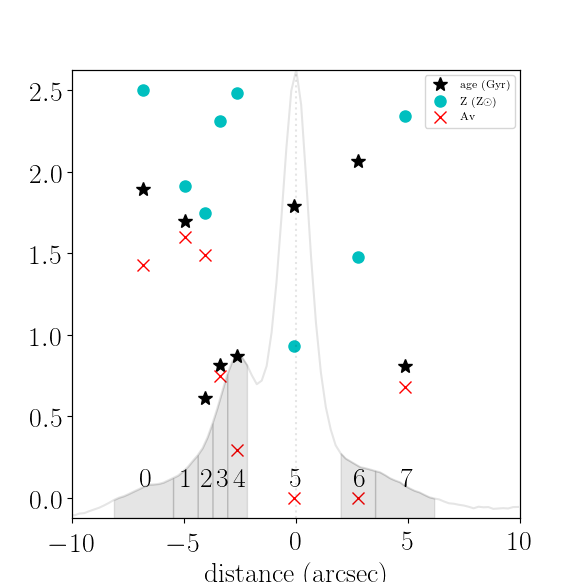} 
\caption{Left: Example fit (masking emission lines) of one of the spectra extracted from the OSIRIS and ISIS long-slit data (black) and the \textsc{starlight} fitted result (red). The Y-axis is in units of $10^{-17}\ \rm erg\ s^{-1}\ cm^{-2}\ \AA$. Right: Results from the stellar population synthesis using \textsc{starlight} applied to the seven spectra extracted from the combined OSIRIS and ISIS long-slit data, as indicated along the slit profile (light gray). The origin of distance in the highest peak corresponds to the \gm-NLSy1, while the lower peak on the left is the Seyfert 2. The Y-axis values apply to the three variables, with the age in Gyr, the metallicity in solar units, and the extinction in magnitudes. There is an inwards age gradient with the youngest ages of $\sim0.5$ Gyr corresponding to the region around the Seyfert 2 nucleus while the outskirts are $\sim2$ Gyr. The metallicities are solar and above across the system, while extinction seems to be larger to the South-East (the halo external to the side of the Seyfert 2, Figure~\ref{subaru}).}\label{spatial_variations}
\end{figure*}

The BPT diagram \citep[][]{1981PASP...93....5B} is a powerful tool to explore the excitation mechanism of the ionized gas. We used the WHT and GTC long-slit spectroscopic observations and extracted spectra at eight locations along the slit  with the motivation to understand the physical behavior of the merging environment. In Figure~\ref{bpt}, we show the location of seven of the eight spectra extracted\footnote{The location of the extracted spectra along the long-slit is shown in the right panel of Figure~\ref{spatial_variations}. The spectrum at location 7 in Figure~\ref{spatial_variations} does not have a measurable H$\beta$ line, hence excluded.} from the long-slit spectroscopy (large red dots) over the background of the BPT diagram of \citet[][]{2006MNRAS.372..961K} for 85224 galaxies from the SDSS data release 4. Because these seven spectra are extracted from the long-slit along a stretch of 15\arcsec~(see black lines in Figure~\ref{sub_J_meg_ha_maps}), we can firmly state that a large fraction of the gas along the slit in the merging system is affected by AGN and shocks, i.e., physical processes other than due to hot, massive stars. The BPT diagram was also used to elucidate the nature of the enhanced H$\alpha$ emitting ring surrounding the \gm-NLSy1 galaxy observed in the MEGARA IFU image (Figure~\ref{sub_J_meg_ha_maps}, top panel) which confirmed that it is excited by the merger induced shocks and the AGN emission.

After correcting for 3.6 \AA\ instrumental resolution, the full width at half maximum (velocity dispersion) value for the emission line [O III] 5007 \AA\ is 10.29 \AA\ (208 km s$^{-1}$) for the Seyfert 2 nucleus. The FWHM[O III] can be converted to the stellar velocity dispersion ($\sigma$) by $\sigma$=FWHM[O~III]/2.35 \citep[][]{1995ApJS...99...67N,2001A&A...377...52W}. Using the known statistical correlation between the mass of the central black hole and $\sigma$ \citep[][]{2001A&A...377...52W,2004ApJ...606L..41G,2009ApJ...698..198G}, this value transforms to a black hole mass of $\sim4\times10^6$ M$\odot$ for the companion Seyfert 2 galaxy. For the \gm-NLSy1 galaxy, on the other hand, we used the well calibrated empirical relations \citep[cf.][]{2011ApJS..194...45S} adopting the conventional H$\beta$ FWHM ($\sim$1900 km s$^{-1}$) and continuum luminosity at 5100 Angstrom \citep[$\sim10^{44}$ erg s$^{-1}$; see also][]{2017ApJS..229...39R}. The derived black hole  mass is $\sim3\times10^7$ M$\odot$, similar to that reported in literature \citep[e.g.,][]{2017ApJS..229...39R}. Note that the typical uncertainties associated with these approaches are 0.3$-$0.5 dex \citep[][]{2006ApJ...641..689V,2009ApJ...698..198G,2011ApJS..194...45S}. Furthermore, since the black hole mass and the host galaxy properties are significantly correlated \citep[][]{2005Natur.433..604D,2015ApJ...813...82R}, the total stellar mass is estimated as $\sim1\times10^{11}$ M$\odot$ and $\sim 2\times10^{10}$ M$\odot$ for the \gm-NLSy1 and the companion galaxy, respectively. Based on the large ratio of the stellar masses, the observed event can be classified as a `minor' merger \citep[][]{2011ApJ...742..103L}.

\subsection{Merger Timescale}
We use the spectral population synthesis code \textsc{starlight} \citep[][]{2005MNRAS.358..363C} to derive key population parameters, such as mean age and metallicity of the merging system. The continuum stellar spectra (masking the emission lines) were fitted for each of the eight spectra extracted from the spatial locations along the OSIRIS and ISIS slits, as demonstrated in the left panel of Figure~\ref{spatial_variations}. We retrieved the star formation history, the dust extinction, mass weighted stellar metallicity, and luminosity weighted stellar ages and show them in the right panel of Figure~\ref{spatial_variations}. We used a set of 163 single stellar populations sampling the 3 Myr to 9.75 Gyr\footnote{Note that the upper limit to the oldest population possible in this system is given by the redshift of 0.26, when the universe was 10.7 Gyr old.} age range with 33 logarithmically spaced age bins and five metallicities between 0.02 and 2.5 solar (S. Charlot and G. Bruzual 2007, private communication), assuming a Chabrier initial mass function. Further details about \textsc{starlight} can be found on the project web page http://www.starlight.ufsc.br, and a review of stellar populations fitting techniques is available at http://www.sedfitting.org \citep[][]{2011Ap&SS.331....1W}.

As can be seen in Figure~\ref{spatial_variations}, the results from the stellar population synthesis calculations performed on the GTC and WHT optical spectra show average (luminosity weighted) stellar ages in the range $\sim$0.5-2.5 Gyr, older in the outer halo and younger in the inner region around the Seyfert 2 nucleus. The presence of the younger stellar population in proximity to the merging nuclei indicates that the starburst activity has recently begun in the central regions. This result is expected if the spectra are dominated by merger driven activity in the past $\sim$0.5-2 Gyr \citep[][]{2015MNRAS.448.1107M}. This is because the star formation history that develops along the evolution of a merging pair of galaxies depends on the distance between the galaxies during the different binding passages. For example, figures 3 and 4 in \citet[][]{2015MNRAS.448.1107M} show the star formation evolution from the approach and first passage until the coalescence. Clearly, the window of higher star formation, i.e., when more gas is available to form stars and also to fall into the nuclei, is in the approximate age range 0.5$-$2 Gyr. The detailed history of star formation and nuclear activity also depends on the relative masses and morphology of the two merging galaxies and on the relative angular momenta of the spirals and the merging orbit; but models show that nuclear activity follows after the main bouts of star formation. Indeed, \citet[][]{2012ApJ...748L...7V} find that strong dual AGN activity occurs during the late phases of the mergers, at small separations ($\lesssim 10$ kpc). Thus, the results from the analysis of the stellar populations and the nuclear activity in TXS 2116$-$077, especially the merger timescale, match quite precisely the expectations from the numerical simulations.

From the total narrow H$\alpha$ flux measured in the MEGARA data (15.7 mJy), the star-formation rate in the system is 3.8 $\rm M_{\odot}\, yr^{-1}$, using the established correlation of star-formation rate with H$\alpha$ emission \citep[][]{1998ARA&A..36..189K}. Using the correlation between the global star-formation rate and the galaxy stellar mass \citep[][]{2016A&A...590A..44G,2017A&A...607A..70C}, this places TXS 2116$-$077 just within but somewhat high in the star-forming main sequence, implying that the star-formation is possibly enhanced by the merging process when compared to normal late-type spiral galaxies \citep[][]{2016A&A...590A..44G}.

\subsection{Kinematic Age of the Radio Jet  and A Comparison with the Merger Timescale}
Since the radio jet in \txs~remains compact down to mili-arcsec scale, we have computed the upper limit to its projected length at $z=0.26$ as 0.02 kpc, using the VLBA beam-size of ($5.3\times2.4$) miliarcsecond$^2$. Considering a viewing angle of 3$^{\circ}$, typical for \gm-ray emitting beamed AGNs \citep[][]{2014Natur.515..376G,2017ApJ...851...33P}, the true length turns out to be $<$1 kpc after correcting for projection effects. This gives the kinematic age of the jet as $<$15 kyr, assuming a typical jet velocity of 0.1c \citep[e.g.,][]{2013A&A...557L..14G}. The age will be even shorter for the larger jet velocity/viewing angle. This finding suggests a `young' jet in \txs, which is in line with the hypothesis that NLSy1 jets are likely to be in the early phase of evolution \citep[][]{2018MNRAS.480.1796S}. Note that the radio emission in NLSy1 galaxies could be originated from the star-formation activities and/or due to non-relativistic kiloparsec-scale outflows \citep[][]{2006AJ....132..546G}. However, the observation of the flux variability, especially in the \gm-ray band, and also the detection of a flat GHz frequency spectrum \citep[][]{2018MNRAS.477.5127Y} indicates that the radio emission detected from TXS 2116$-$077 is jet dominated and hence the derived kinematic age is reliable.

A comparison of the jet kinematic age ($<$15 kyr) with the merger timescale (0.5$-$2 Gyr) suggests the jet to be considerably younger than the merger. This observation provides a supportive evidence for a scenario in which relativistic jets are triggered by galaxy mergers. Though high-quality observations, similar to those presented here, are not available for many jetted AGNs, recent observing campaigns focused on the high-resolution host galaxy imaging of radio-loud NLSy1 galaxies are also converging to this scenario \citep[see, e.g.,][]{2019AJ....157...48B,2020MNRAS.492.1450O}. In particular, \citet[][]{2020MNRAS.492.1450O} studied a sample of \gm-NLSy1s and reported that minor mergers could be responsible for the observed nuclear activity in this class of AGN. Though host galaxy observations of a few other sources revealed a rather secular driven growth of the system \citep[][]{2016ApJ...832..157K} and a few studies also claimed elliptical hosts \citep[e.g.,][]{2017MNRAS.469L..11D}, the majority of the \gm-NLSy1s undoubtedly reside in young, late-type galaxies in which jet launching is likely to be triggered via galaxy mergers. Similar results are found in various studies focusing on more powerful, radio-loud quasars \citep[][]{2013MNRAS.436..997R,2015ApJ...806..147C,2016MNRAS.460.3202O}. The fact that radio-loud NLSy1s are likely to be living in denser large-scale environments \citep[][]{2017A&A...606A...9J} where mergers could be frequent, also provide supportive evidences for the merger triggered jet launching. Combining this with the idea of the young nature of NLSy1s, it can be concluded that \gm-NLSy1 galaxies represent the beginning phase of the jet activity.

There is significant support of the idea that rotating supermassive black holes produce jets and their origin is connected to magnetic fields generated in accretion disks of spinning black holes \citep[][]{1977MNRAS.179..433B,1999ApJ...522..753M}. Recent 3D relativistic magnetohydrodynamic simulations support this spin paradigm \citep[see, e.g.,][]{2011MNRAS.418L..79T} and demonstrate that prograde disk accretion onto rapidly spinning black hole most efficiently generate jets. Combining this with the idea that galaxy mergers trigger AGN \citep[][]{2005Natur.433..604D,2011MNRAS.415.1027H} leads to the expectation that a merger may produce the needed high spin. Galaxy mergers can efficiently drive gas inflows all the way to the central regions of the galaxy and all the way to the accretion disk that feeds the central black hole.

\section{Summary}\label{summary}
We have carried out a comprehensive multi-frequency campaign to study the host galaxy properties of a \gm-NLSy1 galaxy, \txs. The high-resolution imaging done with the Subaru telescope has revealed the \gm-NLSy1 to be merging with a nearby ($\sim$12 kpc) galaxy. Subsequent optical spectroscopic observations from GTC and WHT have confirmed the merging phenomenon and revealed that both galaxies host AGN at their centers with the companion galaxy being a Seyfert 2. Furthermore, IFU observations from MEGARA, along with the long-slit spectroscopy with ISIS at WHT and OSIRIS at GTC, have provided a very rare view of the merging environment surrounding the relativistic jet. We have also found that \txs~hosts a compact radio jet whose kinematic age is considerably smaller than the merger timescale derived from the stellar population synthesis. Our findings are aligned with other recent studies focused on the host galaxy imaging of NLSy1s \citep[][]{2020MNRAS.492.1450O} and support a scenario in which relativistic jets are triggered by galaxy mergers \citep[][]{2015ApJ...806..147C}.

Black holes in NLSy1 galaxies usually grow via secular processes, develop pseudobulges, mostly remain radio-quiet, and evolve to broad-line Seyfert 1s \citep[][]{2000MNRAS.314L..17M}. Those residing in dense large-scale environments can frequently undergo mergers \citep[][]{2015ApJ...806..147C,2017A&A...606A...9J}, triggering rapid black hole growth due to inflow of gas driven by gravitational torques \citep[][]{1991ApJ...370L..65B,1996ApJ...471..115B}, and possibly spin-up the black hole leading to the launch of the relativistic jet. In this evolutionary scenario, \gm-NLSy1s can be considered as nascent blazars and as such they are the best jetted systems to explore the jet launching regions and their associated physical processes.

\bibliography{Master.bib}{}
\bibliographystyle{aasjournal}

\acknowledgments
Thanks are due to the journal referee for a constructive criticism. V.S.P. and M.A. acknowledge the prompt {\it Chandra} DDT observation of TXS 2116$-$077 and funding under contract DD8-19104X. V.S.P. is grateful to C. Y. Peng for his assistance with the GALFIT modeling. M.A. acknowledges discussions with R.~Blandford and C.~Conselice and the support of NSF through grant AST-1715256. F.P. acknowledges support from the Spanish MICINN grant PGC2018-101931-B-100. E.P. and R.G.B. acknowledge financial support from the Spanish Ministry of Economy and Competitiveness through grant AYA2016-77846-P. E.P., R.G.B. and F.P. acknowledge support from the State Agency for Research of the Spanish MCIU through the ``Center of Excellence Severo Ochoa'' award to the Instituto de Astrof\'isica de Andaluc\'ia (SEV-2017-0709). A.D. acknowledges the support of the Ram{\'o}n y Cajal program from the Spanish MINECO. This work is based on data collected at the Subaru Telescope, which is operated by the National Astronomical Observatory of Japan. Based on service observations made with the 4.2m William Herschel Telescope (WHT) operated by the Isaac Newton Group of Telescopes and installed in the Spanish Observatorio del Roque de los Muchachos of the Instituto de Astrofísica de Canarias, in the island of La Palma. We are grateful to staff astronomers Antonio Cabrera at GTC and Ian Skillen at WHT for carrying out MEGARA, OSIRIS and ISIS observations. This work is based on observations made with the GTC telescope, in the Spanish Observatorio del Roque de los Muchachos of the Instituto de Astrofísica de Canarias, under Director’s Discretionary Time.

\vspace{5mm}
\facilities{Subaru, GTC, WHT, \chandra, VLA}

\software{IRAF \citep[][]{1986SPIE..627..733T,1993ASPC...52..173T}, MEGARA data reduction pipeline (S. Pascual 2020, in preparation), AIPS \citep[][]{1996ASPC..101...37V}, CIAO \citep[][]{2006SPIE.6270E..1VF}, XSPEC \citep[][]{Arnaud96}, starlight \citep[][]{2005MNRAS.358..363C}, GALFIT \citep[][]{2010AJ....139.2097P}.}

\end{document}